\documentclass[a4paper,fleqn]{cas-dc}

\usepackage[authoryear,longnamesfirst]{natbib}
\usepackage{adjustbox}
\usepackage[graphicx]{realboxes}
\usepackage{rotating}
\usepackage{adjustbox}
\usepackage[toc,page]{appendix}

\newcommand{\bm}{\boldmath}
\newcommand{\ubm}{\unboldmath}

\def\hlinewd#1{%
\noalign{\ifnum0=`}\fi\hrule \@height #1 %
\futurelet\reserved@a\@xhline} 

\def\tsc#1{\csdef{#1}{\textsc{\lowercase{#1}}\xspace}}
\tsc{WGM}
\tsc{QE}

\begin{document}
\let\WriteBookmarks\relax
\def\floatpagepagefraction{1}
\def\textpagefraction{.001}

\shorttitle{Evidence of a sub-solar star in a microlensing event toward the LMC}    

\shortauthors{A. Franco et al.}  

\title [mode = title]{Evidence of a sub-solar star in a microlensing event toward the LMC}



%

\author[1,2,3]{A. Franco}[type=editor,
       orcid=0000-0002-4761-366X]
\author[1,2,3]{A. A. Nucita}
\author[1,2,3]{F. De~Paolis}
\author[1,2,3]{F. Strafella}

\affiliation[1]{organization={Department of Mathematics and Physics ``E. De Giorgi'' , University of Salento, Via per Arnesano, CP-I93, I-73100, Lecce, Italy}}
\affiliation[2]{organization={INFN, Sezione di Lecce, Via per Arnesano, CP-193, I-73100, Lecce, Italy}}
\affiliation[3]{organization={INAF, Sezione di Lecce, Via per Arnesano, CP-193, I-73100, Lecce, Italy}}

\fntext[1]{Corresponding author: antonio.franco@le.infn.it}


\begin{abstract}
Gravitational microlensing is known to be an impressive tool for searching dark, small, and compact objects that are missed by the usual astronomical observations. 
In this paper, by analysing multiple images acquired by DECam, we present the detection and a complete description of the microlensing event LMC~J05074558-65574990 which is most likely due to a sub-solar object with mass $(0.16\pm0.10)$~M$_\odot$, hence in the mass range between a massive brown dwarf and a red dwarf, whose distance is estimated to be $7.8^{+4.1}_{-3.4}\times10^2$~pc thanks to the Gaia observation of the source, leading us to consider this lens as one the closest ever detected.

\end{abstract}



\begin{keywords}
Physical Data and Processes: gravitational lensing; methods: data analysis; techniques: image processing; (galaxies:) Large Magellanic Clouds.
\end{keywords}
\maketitle

\section{Introduction}

Since the discovery of the first microlensing event in 1989 (\citealt{irwin}) and the results obtained by some important observational campaigns like OGLE (\citealt{ogle_first}), MACHO (\citealt{macho_first}) and EROS (\citealt{eros_first}), the search for microlensing candidates became more resolute. In particular, dense and crowded regions, such as those toward the Galactic Bulge and the Magellanic Clouds, have always been preferred in order to test new galactic models, searching for any evidence of dark objects acting as gravitational lenses.

For events induced by a single lens, the typical brightness profile, the so called Paczy\'nski function (\citealt{paczynski86}), shows a flux variation delineated by a symmetric and achromatic light curve. However, sometimes the light curve may be distorted or broadened due to secondary effects, such as the Earth parallax (due to annual Earth motion around the Sun) and the finite source (due to the not negligible finite angular size of the source that cannot be approximated to a single point) as described in \citet{gould92} and \citet{gould00}. 

By fitting a simple Point-Lens Point-Source event, one cannot entirely resolve the system, since there is no way to obtain a direct estimate of both the lens distance and mass. However a Monte Carlo simulation can give useful statistical indication of the system parameters (see \citealt{ingrosso2006}).

In the present paper we report about the discovery of a microlensing event labelled as LMC~J05074558-65574990 ($\alpha=05^h07^m45.58^s$, $\delta=-65^\circ57'49.90"$; $l=276^\circ.662903$, $b=-36^\circ.218246$) as a possible sub-solar object candidate, particularly close to Earth. We show the result obtained by analysing repeated images acquired by DECam in three SDSS bands ({\it g}, {\it r}, and {\it i}) confirming the achromaticity feature of the detected microlensing event, following the same data reduction and photometric procedure adopted in \citet{franco}.

The paper is structured as follows: in Section \ref{Sec:obs}, we give some details about the observational survey considered in this work, some information about the target source and a description of the analysis conducted in order to obtain the calibrated light curves. We also describe the usage of the the {\it pyLIMA} software (\citealt{pylima}) for the modelling of the microlensing event. In section \ref{Sec:micromodel}, we give details on the different microlensing models taken into account specifying the best one we chose, i.e. the simpler Point-Source Point-Lens model, addressing then our conclusion in Section \ref{Sec:results}.

\section{Observations and data reduction}
\label{Sec:obs}

The discovered microlensing event has been detected by analysing multiple temporally sorted images acquired by DECam (Dark Energy Camera)\footnote{The Dark Energy Camera is a high-performance, wide-field camera installed at the 4m V. Blanco Telescope, CTIO, Chile (\citealt{flaugher}).} during the two-year CTIO program (2018A-0273)\footnote{The proposed survey project with the title ``PALS: Paralensing Survey of Intermediate Mass Black Holes'' was originally planned to potentially identify microlensing events induced by Intermediate-Mass Black Holes.}, intensively observing the Magellanic Clouds between February 2018 and January 2020 (see \citet{franco2021} and \citet{franco} for more details). In particular, we used observations obtained in three different DECam bands, i.e. {\it g}, {\it r}, and {\it i}, that correspond to the respective SDSS bands. Specifically, the observed source has been captured in 37 {\it g}~band, 49 {\it r}~band, and 7 {\it i}~band images. We also checked {\it ESA XMM-Newton} and {\it NASA Swift} images for any X-ray counterpart of the target source, considering its potential as a cataclysmic variable. However, we did not find any correlation, reinforcing the conclusion that the most plausible explanation remains the microlensing scenario.

\begin{figure*}
    \centering
    \includegraphics[width=0.7\textwidth]{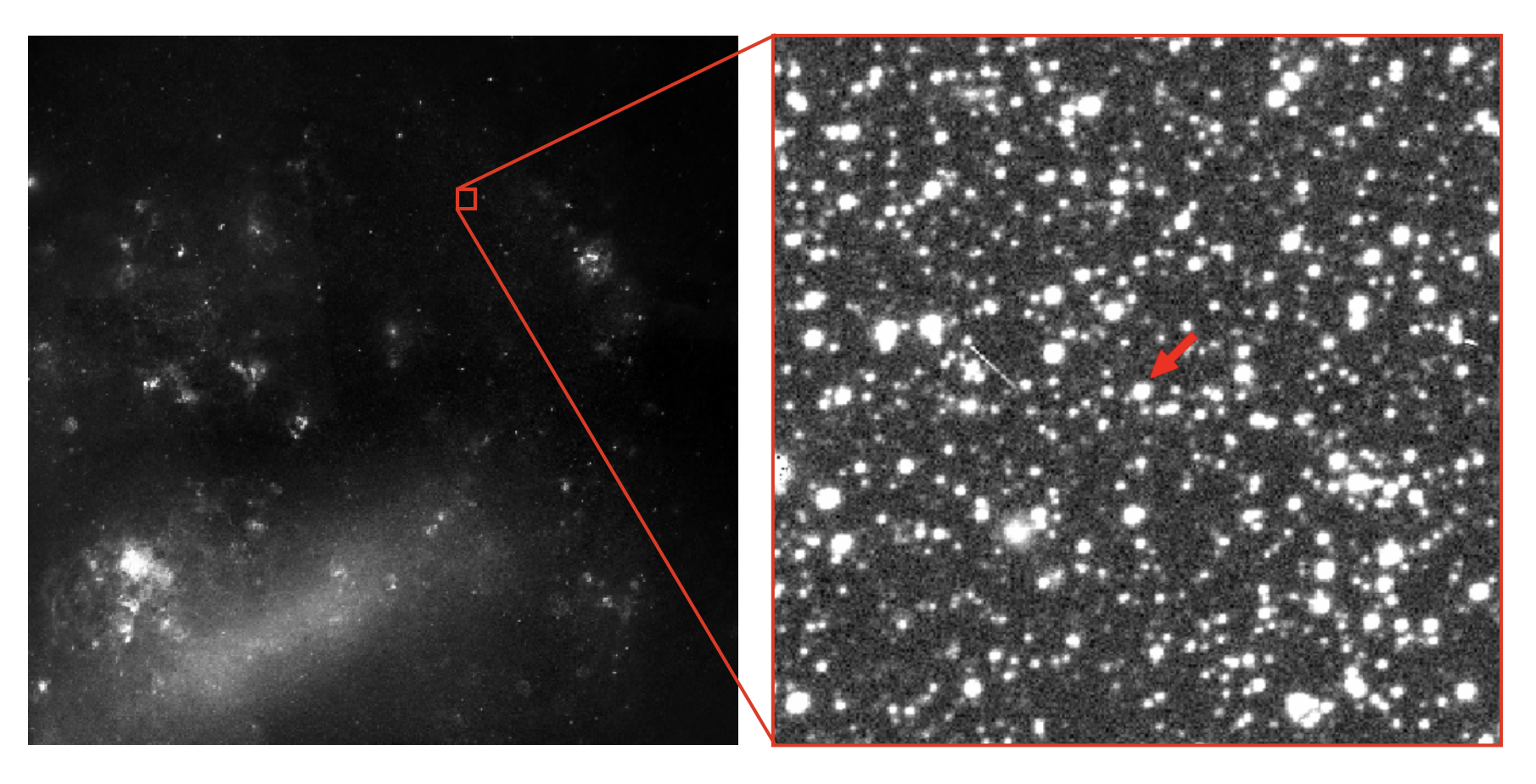}
    \caption{Large portion of the LMC (left panel) in which the interested DECam field is observed (right panel). The red arrow in the right panel indicates the microlensing candidate position as observed by DECam.}
    \label{Fig:lmc_field}
\end{figure*}

In Figure \ref{Fig:lmc_field} we present an image of the LMC field, obtained by DECam, in which the lensed source star is located.
The star is also present in GaiaDR3 (\citealt{gaiadr3}), where it is catalogued as a star located in the Galactic thick-disk at a distance of $1.55^{+0.72}_{-0.53}$ kpc from Earth (\citealt{dist_gaiaedr3}) that derives from the assumption that the object is single (as explicitly remarked by \citealt{dist_gaiaedr3}) \footnote{We note that the source distance is not the typical one expected for the sources in a microlensing campaign toward the LMC.}.
The Gaia magnitudes reported in the catalogue are 20.33, 20.88, and 19.77 in the G, BP, and RP bands, respectively. 

The source surface temperature can be estimated by considering the B-V colour index. The observed values can be extracted from the catalogue ``Magellanic Clouds Photometric Survey: the LMC'' (\citealt{zaritsky}), which returns $B-V=0.936\pm0.130$. By querying the ADS/IRSA database\footnote{Data Tag: ADS/IRSA.Dust\#2023/0719/032229\_22258} at the object coordinates, the respective colour excess provided by \citet{schlafly} is $E(B-V)=0.0812 \pm 0.0049$, and the true colour index becomes $B-V=0.855\pm0.130$. Considering the well known colour index relation with the associated temperature, given by Ballesteros' formula (\citealt{temperature}), the  obtained surface temperature of the lensed star turns out to be $T_S=(5120 \pm 370)\>{\rm K}$, corresponding to a K1V spectral type star (\citealt{lang}).

With the aim of obtaining the light curve of the source star, the acquired DECam observations, already corrected by bias, flat, and dark, have been aligned and processed by using the ISIS 2.2 subtraction package (\citealt{alard98, alard00}), a powerful software capable of highlighting variable sources in a set of multiple temporally sorted images, employing a photometric analysis based on the difference between homogeneous images. The software returns a list of differential fluxes, evaluated as
\begin{equation}
    \Delta f_{\rm i} = f_{\rm ref}-f_{\rm i}
\end{equation}
where $f_{\rm ref}$ is the flux related to a reference image, chosen as that being characterised by the best signal-to-noise ratio, and $f_i$ represents the flux corresponding to the i-th image of the sample. Therefore, by evaluating a source flux ($f_{\rm ref}$) in the reference image with the DAOPHOT/ALLSTAR softwares (\citealt{daophot}), the flux of the same source in the i-th image is simply obtained by inverting the previous equation, thus giving 
\begin{equation}
    f_{\rm i} = f_{\rm ref}-\Delta f_{\rm i}
\end{equation}
By adopting the procedure described in \citet{franco}, we calibrated photometrically all the images with the ATLAS All-Sky Stellar Reference Catalog (\citealt{tonry}) and obtained the calibrated DECam magnitudes in the $g$, $r$, and $i$ bands as
\begin{equation}
    \begin{split}
        g &= 0.928 \cdot g_{\rm instr} + 6.463 \\
        r &= 0.932 \cdot r_{\rm instr} + 6.082 \\
        i &= 1.105 \cdot i_{\rm instr} + 4.902
    \end{split}
\end{equation}
where {\it $g_{\rm instr}$}, {\it $r_{\rm instr}$}, and {\it $i_{\rm instr}$} are the corresponding instrumental magnitudes. 

Once the photometric procedure was applied, the analysis of the microlensing candidate has been carried out with {\it pyLIMA}, a Python open-source package for microlensing modelling (\citealt{pylima}). We considered both a Point-Lens Point-Source (PSPL) model and a Finite-Source Point-Lens (FSPL) model,  accounting also for the annual Earth parallax effect, as described in \citet{gould92} and \citet{alcock95}.

Based on this model, the fit considers up to twelve parameters, depending on the microlensing model considered: the time $t_0$ of minimum approach between the source and the lens, the minimum impact parameter $u_0$, the Einstein time $t_{\rm E}$, the source finite angular radius $\rho$ in units of Einstein angle, the North and East parallax components $\pi_{\rm EN}$ and $\pi_{\rm EE}$, the baseline fluxes in the three bands (from which one can estimate the baseline magnitudes $m_{\rm 0,g}$, $m_{\rm 0,r}$ and $m_{\rm 0,i}$) as well as the blending factors $g_{\rm g}$, $g_{\rm r}$ and $g_{\rm i}$ (see Sec. \ref{Sec:micromodel} for details).

\begin{figure*}
    \centering
    \includegraphics[width=0.45\textwidth]{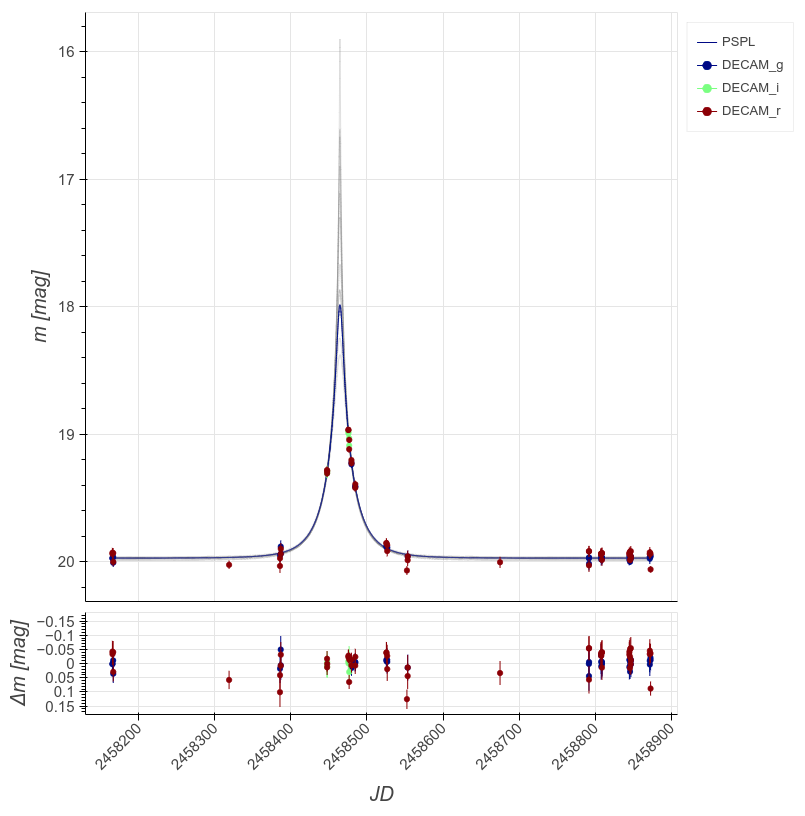}
    \includegraphics[width=0.45\textwidth]{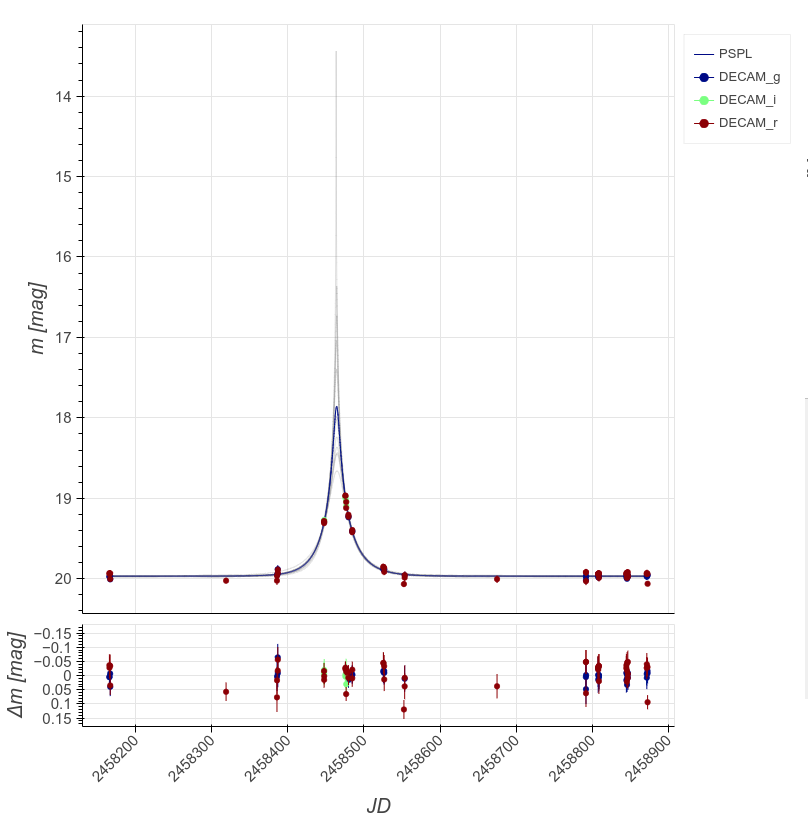}
    \includegraphics[width=0.45\textwidth]{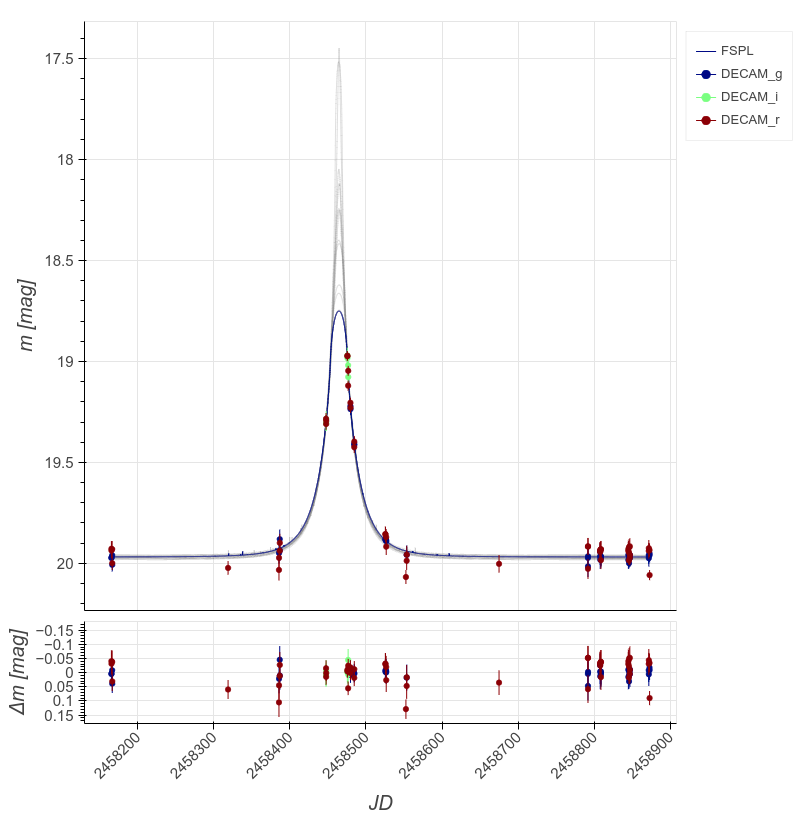}
    \includegraphics[width=0.45\textwidth]{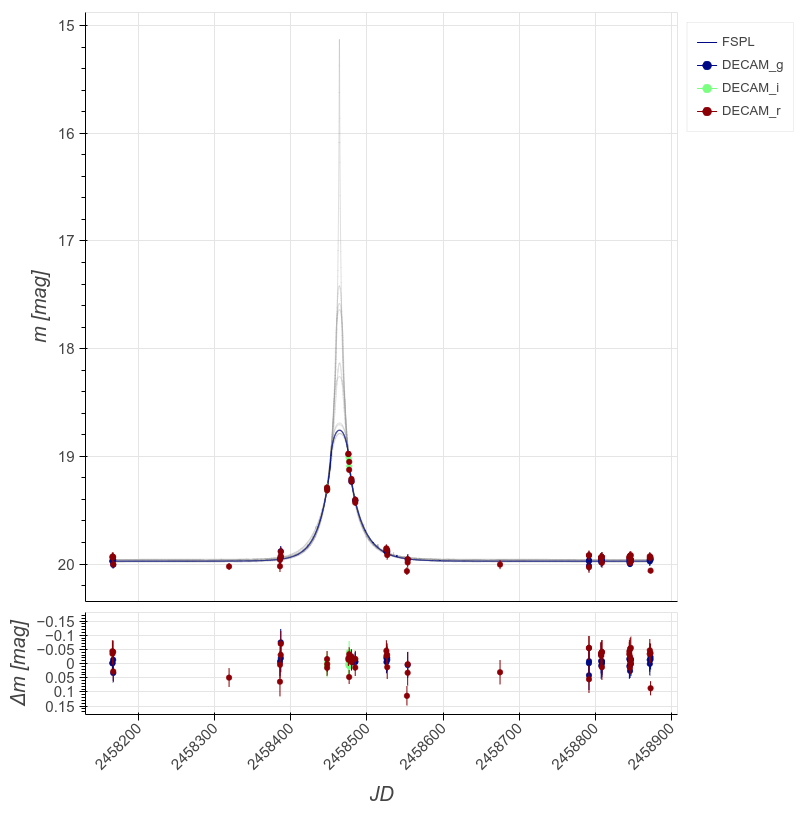}
    \caption{In the four panels, the g-band (blue dots), r-band (red dots), and i-band (green dots) DECam data for the microlensing candidate are reported. The blue curve represents the best fit curve corresponding to the best-fit model adopted. In particular, the considered model corresponds to a Point-Source Point-Lens model (upper left), Point-Source Point-Lens model + Earth Parallax effect (upper right), Finite-Source Point-Lens (lower left) and Finite-Source Point-Lens + Earth Parallax (lower right). For each panel, the subpanel in the bottom shows the residuals for each band.}
    \label{Fig:fit}
\end{figure*}

\section{Microlensing model}
\label{Sec:micromodel}

\begin{figure}
    \includegraphics[width=\columnwidth]{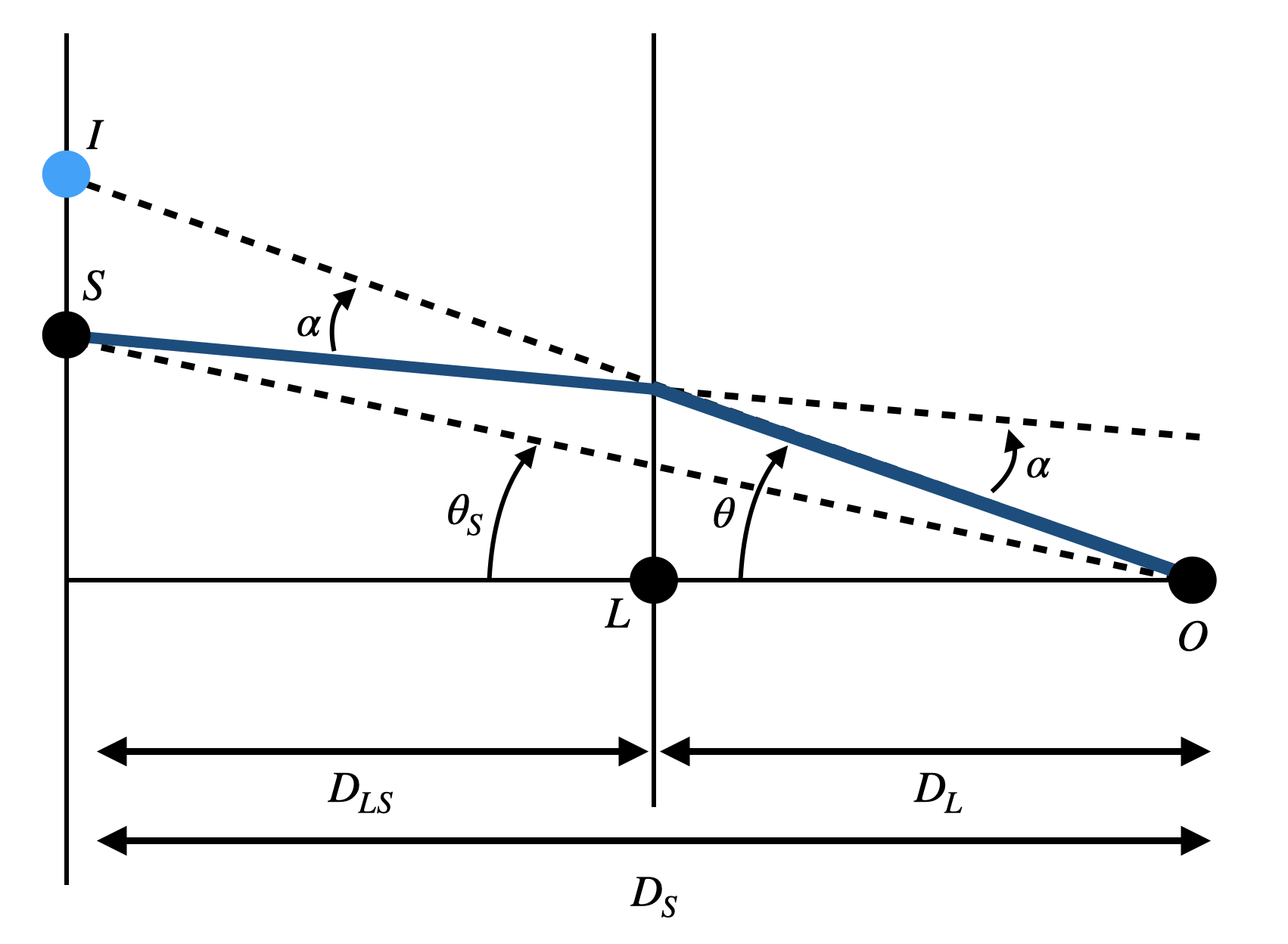}
    \caption{Gravitational lensing geometry (see, e.g., \citealt{depaolis}). The black dots represent the source (S), lens (L), and observer (O), respectively. The blue dot, labelled as I, indicates the position of the virtual image produced by the light  deflection. The blue line indicates the path followed by light. Note that $D_{\rm L}$, $D_{\rm S}$, and $D_{\rm LS}$ are the distances to the lens, to the source star, and the lens-source distance, respectively. The angle $\alpha$ indicates the deflection angle, $\theta_S$ is the lens-source angular distance and $\theta$ is the lens-image angular distance.}
    \label{Fig:lens_geo}
\end{figure}

Gravitational lensing, i.e. one of the natural consequences of Einstein's General Relativity,  shows up thanks to the gravitational effect of a massive object, the lens, which bends the light rays from a distant source, generating multiple virtual images, whose deflection angle (see Figure \ref{Fig:lens_geo}) is $\alpha \approx \frac{4GM_L}{\theta D_L c^2}$ (\citealt{einstein1936}), where $M_L$ is the lensing object mass and $D_L$ is its distance from Earth. 

In a microlensing event, the angular separation of the produced images cannot be spatially resolved by a telescope. However, a light magnification $A(t)$ of the source brightness (\citealt{paczynski86, paczynski96}) given by 
\begin{equation}
\label{Eq:ampl}
    A(t)=\frac{u(t)^2+2}{u\sqrt{u(t)^2+4}}
\end{equation}
can be observable. Here, $u(t)=\theta_S(t)/\theta_{\rm E}$ is the source-lens angular separation in Einstein angle unit (Figure \ref{Fig:lens_geo}). It can be also expressed in terms of the event time-scale, named Einstein time $t_{\rm E}$
\begin{equation}
    u(t)=\sqrt{u_0^2+\bigg(\frac{t-t_0}{t_{\rm E}}\bigg)^2}
\end{equation}
where, $u_0$ and $t_0$ are the impact parameter and the time, respectively, both considered at the minimum lens-photon approach.

The Einstein time depends on the transverse lens velocity relative to the observer-source line of sight $v_t$, and the Einstein radius $R_{\rm E}$ as
\begin{equation}
    \label{Eq:etime}
    t_{\rm E}=\frac{R_{\rm E}}{v_t} = \frac{\theta_{\rm E} D_{\rm L}}{v_{\rm t}}
\end{equation}
where $\theta_E=\sqrt{\frac{4GM_L}{c^2}\frac{D_S-D_L}{D_S D_L}}$ is the Einstein angle.

The amplified magnitude of the lensed stars can be computed through the usual relation
\begin{equation}
    m(t) = m_0 - 2.5 \log_{10} [(A(t)-1)g+1] - 2.5 \log_{10} g
\end{equation}
where $m_0$ is the baseline magnitude  and $g$ is the blending parameter. The latter parameter accounts for any possible flux contribution of the lensing object and/or nearby stars and is defined as
\begin{equation}
    g = \frac{f_{\rm b}}{f_{\rm s}}
\end{equation}
where $f_{\rm s}$ is the source flux\footnote{Since the fit procedure returns the baseline flux in each band, we converted the fluxes into magnitudes by considering the {\it pyLIMA} zeropoint, i.e. $m_{\rm ZP}=27.4$, and the relation $m_0=m_{\rm ZP}-2.5\cdot\log_{10}f_{s}$ (\citealt{pylima})} and $f_{\rm b}$ is the blended flux.

In this work, we considered four different models in order to better explain the analysed data. In particular, we firstly fit our data with a simple Point-Lens Point-Source (PSPL) model, in order to obtain a preliminary set of reliable values and good boundaries for the involved parameters. Afterwards, we related more accurate models: in particular, considering the Earth Parallax and the finite source possibility, we extended the first model adopted by using the PSPL+Earth Parallax and the Finite-Source Point-Lens (FSPL) models. Lastly, the most complex model, namely FSPL+Earth Parallax, has been taken into account. In the scenario of an extended source, the total amplification is obtained by integrating the amplification in Eq. \eqref{Eq:ampl} over the source area $S$. Considering the source angular radius ($\theta_S$) in units of Einstein's angle $\rho=\theta_S/\theta_E$, the total amplification is
\begin{equation}
    A^*(u,\rho) = \frac{\int_S A(t)\> dS}{\int_S dS}=\frac{1}{\pi \rho^2}\int_S A(t)\> dS
\end{equation}
where $A(t)$ is given by the Point-Source Point-Lens model in Eq. (\ref{Eq:ampl}).
In the hypothesis of finite source effect, since our data show a quite small magnitude magnification of $\simeq 1\div2$, we are not able to introduce the approximated formula for $A(t)$ as proposed by \citet{gould94}\footnote{For the sake of completeness, if the lens-source separation is quite small (i.e. $u\ll1$), following the Gould approximation (\citealt{gould94}), the Point-Source amplification in Eq. (\ref{Eq:ampl}) can be approximated as $A(t)\approx\frac{1}{u}$ and the amplification in the finite-source model can be obtained by solving elliptic integrals (for further details see \citealt{witt1994, yoo04, cassan06, lee09}).}. Accordingly, we took into account the full non-approximated expression, implemented in pyLIMA by following a robust algorithm proposed by \citet{lee09}.

Moreover, since $\rho$ is defined as the source size in units of the Einstein angle, estimating the $\rho$ value makes easier estimating  the Einstein angle through the relation
\begin{equation}
\label{Eq:thetaE}
    \theta_{\rm E} = \frac{R_S}{\rho D_S}
\end{equation}
where $R_S$ and $D_S$ are the source radius and the Earth-source distance, respectively.

Another relevant information that one can extract from the fit procedure is that related to the parallax effect induced by the Earth's motion around the Sun. It allows to estimate the value of the Einstein radius projected onto the observer plane, i.e. the reduced Einstein radius $\Tilde{r}_{\rm E}$ (\citealt{gould92, gould00}) given by
\begin{equation}
    \Tilde{r}_{\rm E} = \theta_{\rm E} \> \frac{D_{\rm L}D_{\rm S}}{D_{\rm LS}}
\end{equation}
The parallax parameter is then related to the reduced Einstein radius and to the Earth-Sun distance through the relation
\begin{equation}
    \pi_{\rm E} = \frac{\text{1 AU}}{\Tilde{r}_{\rm E}}
\end{equation} 
where 1 AU$=1.496~\times~10^{8}$~km indicates the Astronomical Unit value. 

The standard parametrization in terms of the microlensing parallax vector components takes into account the East and North components, i.e. $\pi_{\rm EE}$ and $\pi_{\rm EN}$ respectively, whose composition gives the $\pi_{\rm E}$ vector absolute values
\begin{equation}
    \pi_{\rm E} = \sqrt{\pi_{\rm EE}^2 + \pi_{\rm EN}^2}
\end{equation}

By estimating the values of both the parallax parameter  and the Einstein angle, one can combine the equations to obtain a relation for the lens mass (\citealt{gould00})
\begin{equation}
\label{Eq:lensmass}
    M_L=\frac{\theta_{\rm E}}{\kappa \pi_{\rm E}}
\end{equation}
where $\kappa \equiv \frac{4G}{c^2 \text{ AU}} = \frac{3v^2_\oplus}{M_\odot c^2} \simeq 8.144 \> \frac{\text{mas}}{M_\odot}$ and $v_\oplus\simeq 30$ km s$^{-1}$ is the Earth orbital speed.

Finally, considering  Eq. (\ref{Eq:etime}), one can also estimate the lens velocity, obtaining in this way some hints on the possible lens nature.

\begin{table*}
    \caption{Best fit results for the parameters of the four models considered for the LMC~J05074558-65574990 microlensing event light curve. Asterisks (*) indicate the parameters not requested in the considered model.}
    \label{Table:parameters_fit}
    \centering
    \begin{tabular}{lcccc} 
        \hline \hline
        \textbf{ } & \textbf{PSPL}  &   \textbf{PSPL + Parallax} & \textbf{FSPL} & \textbf{FSPL + Parallax} \\
        \hline
        {\bm $t_0$ \ubm {\scriptsize[MJD]}} & $58464.6\pm 0.13$ & $58355.4 \pm 72.5$ & $58464.6 \pm 0.20$ & $58323.18 \pm 52.8$ \\
        {\bm $u_0$ \ubm} & $0.10 \pm 0.07$ & $0.85 \pm 1.23$ & $0.32 \pm 0.09$ & $0.52 \pm 1.52$ \\
        {\bm $t_{\rm E}$ \ubm {\scriptsize[days]}}  & $40.2 \pm 3.0$ &  $38.7 \pm 1.5$ & $38.2 \pm 2.8$ & $33.8 \pm 1.2$ \\  
        {\bm $\rho$ \ubm} & * & * & $0.44 \pm 0.16$ & $0.52 \pm 0.17$  \\
        {\bm $\pi_{\rm EN}$ \ubm} & * & $-0.15 \pm 0.20$ & * & $-0.28\pm0.22$ \\
        {\bm $\pi_{\rm EE}$ \ubm} & * & $0.56 \pm 0.30$ & * & $0.60 \pm 0.25$ \\
        {\bm $m_{\rm 0, g}$ \ubm} & $20.55 \pm 0.16$ &  $20.48 \pm 0.10$  &  $20.37 \pm 0.13$ & $20.25 \pm 0.10$ \\
        {\bm $g_{\rm g}$ \ubm} &  $0.71 \pm 0.22$ & $0.61 \pm 0.14$ & $0.45 \pm 0.20$ & $0.18 \pm 0.09$ \\
        {\bm $m_{\rm 0, i}$ \ubm} & $20.15 \pm 0.23$ & $20.08 \pm 0.10$ & $19.96 \pm 0.13$  & $19.879 \pm 0.09$ \\
        {\bm $g_{\rm i}$ \ubm} & $0.03 \pm 0.31$ &  $-0.19 \pm 0.23$ & $-0.04 \pm 0.28$ & $-0.31 \pm 0.19$ \\
        {\bm $m_{\rm 0, r}$ \ubm} & $20.05 \pm 0.20$ & $19.92 \pm 0.12$ & $19.92 \pm 0.17$ & $19.68 \pm 0.14$ \\
        {\bm $g_{\rm r}$ \ubm} & $0.02 \pm 0.13$ & $-0.04 \pm 0.08$ & $-0.12 \pm 0.14$ & $-0.30 \pm 0.05$ \\
        {\bm  $\chi^2$ \ubm {\scriptsize [NO-SCALING]}} & 192 & 186 & 180 & 175 \\
        {\bm $\chi^2/dof$ \ubm {\scriptsize [NO-SCALING]}} & 2.29 & 2.27 & 2.17 & 2.15 \\
        {\bm  $\chi^2$ \ubm} & 85 & 83 & 83 & 78 \\
        {\bm $\chi^2/dof$ \ubm} & 1.0 & 1.0 & 1.0 & 0.96 \\
        \hline \hline
    \end{tabular}
\end{table*}

\section{Results and Discussion}
\label{Sec:results}
In this paper we presented the discovery of the microlensing event LMC~J05074558-65574990 and the results of the performed analysis. Observations show that the source star is quite close to the Earth, being at a distance of about 1.5 kpc. 

We model the microlensing event in four different ways: first of all, we consider the simpler Point-Source Point-Lens model, which refers to a simple Paczy\'nski light curve. After that, we separately add the parallax and the finite source effect (FSPL) and, at the end, we consider the Finite-Source Point-Lens model taking also into account the Earth parallax effect. The resulted fits for each adopted model mentioned previously are shown in Figure \ref{Fig:fit}, at the top of each sub-panel: the panels on the top refer to the PSPL (left) and PSPL+Earth Parallax (right) models; the panels on the bottom refer to the FSPL (left) and FSPL+Earth Parallax (right) models. We note that the blue line refers to the best fit of the adopted microlensing model, regarding the observed data in the three bands (g, r and i band). The lower sub-panels report instead the residual between the data and the model. The related best fit parameters are presented in Table \ref{Table:parameters_fit}, associated to the corresponding model taken into account. The error bars shown in Figure \ref{Fig:fit} are obtained by multiplying the uncertainties by a constant, defined as the square root of the $\chi^2/dof$ value, for each model. The uncertainties are than calculated as 
\begin{equation}
    \sigma_{\rm mag, i} = \sigma_{\rm mag, i} 
    \times \sqrt{\left(\chi^2/dof\right)_{\footnotesize \rm i}}
\end{equation}
where {\it i} denotes each involved model. Since the $\left(\chi^2/dof\right)_{\footnotesize \rm i}$ obtained values turn out to be quite similar for the different models, one has $\sqrt{\left(\chi^2/dof\right)_{\footnotesize \rm i}}\simeq1.45$. The $\chi^2/dof$ values are then scaled to one. In Table \ref{Table:parameters_fit}, the $\chi^2$ and $\chi^2/dof$ values, obtained before and after the scaling, are reported for completeness.
Moreover, in Figures \ref{Fig:fit_distr1}--\ref{Fig:fit_distr4} reported in Appendix \ref{Sec:appendix1} we present the corner plots with the probability distribution of the parameters involved in the microlensing models. Each column corresponds to a parameter, following the same order as in Table \ref{Table:parameters_fit} for the corresponding model. The one-dimensional histograms are associated to the estimated Monte Carlo probability distribution of each model parameter, while the other plots show the two-dimensional histograms that correlate each pair of parameters. 
\begin{figure*}[ht]
    \centering
    \includegraphics[width=0.45\textwidth]{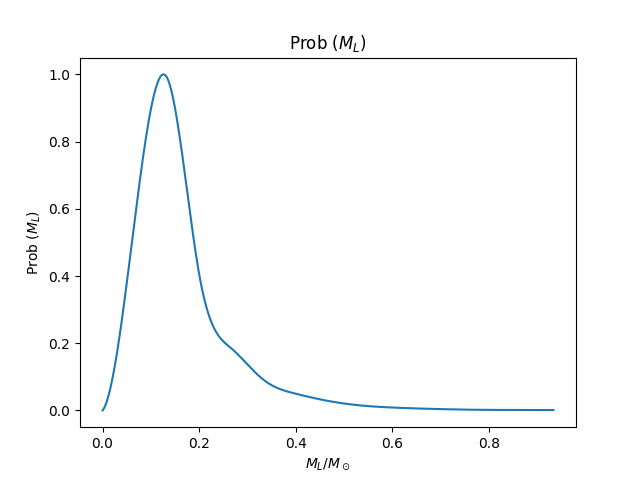}
    \includegraphics[width=0.45\textwidth]{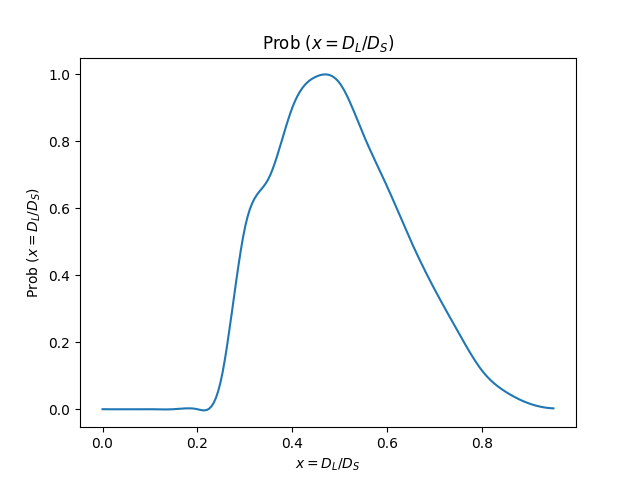}
    \caption{Probability distributions, obtained from the Monte Carlo procedure described in \citet{ingrosso2006}, for the lens mass $M_L$ in units of solar mass (left panel) and the dimensionless lens distance $x$ (right panel) considering simulated events that satisfy the observational parameters.}
    \label{Fig:motecarlo_ml}
\end{figure*}
We would like to point out that we also considered the possibility to fit our data with a binary lens model. Nevertheless, since the data sampling is not very regular and sufficiently dense as one would need to characterise the possible binary systems features, we decided to rely only on the presented models.

We took into consideration all the competitive scenarios in order to account for the light curve of the detected microlensing events. First of all we accounted for the possibility that the Gaia star, lying in the thick-disk of the Milky Way, represents the microlensing lens, while the source is an unresolved star located either in the Galactic halo or in the LMC. 
With this configuration, considering the lens and source distances and the lens mass already reported above, estimating the Einstein angle and using Eq. \eqref{Eq:etime} we get a lens crossing velocity $\gtrsim 100$~km/s. Since one would expect a lens speed roughly close to the typical velocities for thick-disk objects, i.e. $\sim 30\div50$~km/s, the obtained result seems to be rather weak. 
Despite the interesting configuration of this scenario, the blending parameter estimated by the microlensing analysis is compatible with zero, thus confirming that the total flux is dominated by the source, the lens nature of the Gaia star seems quite unlikely.

At this point, the most probable configuration considers the Gaia star as the source star of the  detected microlensing event, while the lens is some fainter object located between the Earth and the source star. We remind that the candidate lensed source has been catalogued as a star by GaiaDR3 (\citealt{gaiadr3}) and by the Magellanic Clouds Photometric Survey (\citealt{zaritsky}) located in the Galactic thick-disk at distance $D_S = 1.55^{+0.72}_{-0.53}$ kpc (\citealt{dist_gaiaedr3}). As introduced in Sec. \ref{Sec:obs}, the colour-index of the star gives an estimate of the effective temperature, $T_S\simeq 5120$~K, that corresponds to a K1V spectral type object, i.e. a red main sequence star, with a typical radius of $R_S \sim 0.797~\>R_\odot$ and mass of $M_S \sim 0.86~\>M_\odot$(\citealt{pecaut, mamajek}).

Considering the microlensing models adopted and the fit results in Table \ref{Table:parameters_fit}, a first look to the $\chi^2$ values might suggest that the FSPL + Earth Parallax model is the best one, even if all other models still show a similar $\chi^2$ values. However, even though more accurate and sophisticated models have been suggested, it appears that the simplest model, i.e. the PSPL model, is the most effective and reliable. Nevertheless, for the sake of discussion, with the assumption that the FSPL + Earth Parallax is chosen as the potential model, the estimate of the lens mass and distance following Eqs. \eqref{Eq:lensmass} and \eqref{Eq:etime} might suggest a microlensing lens with mass $M_L=33^{+17}_{-13}\>M_\oplus$ located at $D_L=1.50^{+0.66}_{-0.49}\>$~kpc from Earth, suggesting a possible free-floating planet nature for the lens.

With these considerations, we assume the PSPL scenario as the preferred one. This model relies on well-defined and restricted estimates for all the involved parameters. Furthermore, the $u_0$ values has a more bounded estimate for the first and third models in Table \ref{Table:parameters_fit}: in fact, since the microlensing amplification is related to $u_0$ as shown in Eq. \eqref{Eq:ampl}, in the FSPL + Earth Parallax model the amplification should be of the order of $10^2$, which involves in the light curve $\Delta {\rm m} \simeq 5\div6$, far from the observed value. In all other cases, the $u_0$ estimates lead to a $\Delta {\rm m} \simeq 2 - 3$, which is instead in line with what is observed in the light curves.
With reference to the fit parameters, reported in Table \ref{Table:parameters_fit}, we can estimate the most probable lens distance $D_L$ and mass $M_L$ by using a Monte Carlo procedure (\citealt{yee2012}). We modelled the Milky Way with a triaxial bulge (\citealt{dwek1995}) and a double exponential stellar disk (\citealt{bahcall1983, gilmore1989}) and consequently we evaluated the expected microlensing event rate $\Gamma (D_L, D_S, M_L, v_t, \mathscr{M})$ as described in \citet{ingrosso2006}, where  $\mathscr{M}$ represents the source absolute magnitude. After defining $x = D_L/D_S$ as the dimensionless lens distance and evaluating the differential rates $d\Gamma/dx$ and $d\Gamma/dM_L$ by integrating over all the remaining relevant quantities, the most likely values of the lens distance and mass can be evaluated as
\begin{equation}
    \overline{x}=\int x\left(\frac{d\Gamma}{dx}\right)dx
\end{equation}
\begin{equation}
    \overline{M}_L=\int M_L\left(\frac{d\Gamma}{dM_L}\right)dM_L
\end{equation}
In this way we obtained an estimate of the average dimensionless lens distance $\overline{x}=0.50\pm0.13$ and average lens mass $\overline{M}_L=(0.16\pm0.10)\>M_\odot$ (see panels in Figure \ref{Fig:motecarlo_ml}). By converting the $\overline{x}$ value in physical units we obtain the lens distance $D_L=7.8^{+4.1}_{-3.4}\times10^2$~pc. In this context, the detected lens can be consider as one of the closest ever detected. 

We also pointed out the possibility of estimating the lens mass by photometric considerations. In the PSPL scenario, by considering the results returned by the fit, one obtains that the g-band and r-band magnitudes for the lens are $g_L\gtrsim21.9$ and $r_L\gtrsim20.1$ due to the blending factors. By converting the magnitudes from Sloan to the V-Johnson, following \citet{jordi2006}, we obtain $V_L\gtrsim21$. Considering that the lens distance is surely smaller then the source distance, i.e. $D_L < D_S\simeq1.55$~kpc, from the definition of the distance modulus we can extract the lower limit of the absolute V magnitude, that comes out to be $V_L^{abs}\gtrsim10$. From \citet{mamajek}, we determine the maximum lens mass, for a main-sequence star, i.e. $M_L\lesssim 0.25 M_\odot$, which agrees with the values obtained from the Monte Carlo.

On the other and, we can also provide an estimate for the lens distance by assuming the lens mass provided by the Monte Carlo, i.e. $M_L\lesssim 0.26 M_\odot$. By considering the previous estimate of the V magnitude $V_L\gtrsim21$, and assuming an absolute magnitude $M_V\gtrsim 12.5$ (as reported in \citealt{mamajek}), from the distance modulus we can estimate a lens distance $D_L\gtrsim 500$ pc that, even is this case, agrees with the Monte Carlo estimate.

The event presented here may also provide further motivation for the study of dark matter and primordial objects possibly formed in the initial stage of the universe and nowadays distributed in the thick Galactic disk. Past, current and, particularly, future surveys, including OGLE (\citealt{ogle_first}), Gaia (\citealt{gaia2016}), and the Vera Rubin Observatory (\citealt{ivezic2019}), are crucial in mapping the Galaxy looking for hidden dark microlensing objects, as well as for the study of stellar populations in the thick Galactic disk, whose recent investigation shows evidence of an older stellar population (see, e.g., \citealt{carollo2019}).


\section*{Acknowledgements}

This paper is based on publicly available observations by DECam (Dark Energy Camera), an instrument mounted on the V. Blanco Telescope, as part of the Cerro Tololo Inter-American Observatory (Chile). DECam images used for this work are publicly available at the \url{https://astroarchive.noirlab.edu/portal/search/\#/search-form} webpage.

\noindent We warmly thank Prof. Gabriele Ingrosso for useful suggestions and advice and for friendly discussions. We also thank for partial support the INFN projects TAsP and Euclid.


\appendix
\onecolumn 

\section{Corner plots}
\label{Sec:appendix1}

In this section are reported the corner plots showing the correlation among the values of the model parameters whose sequence follows that in Table \ref{Table:parameters_fit} for each presented model: horizontal axis from left to right, vertical axis from top to bottom. On the top of each column the one-dimensional distribution of the corresponding parameter is shown, while the two-dimensional plots report the correlation between each pair of parameters.

\begin{figure}[h]
    \includegraphics[width=1.0\textwidth]{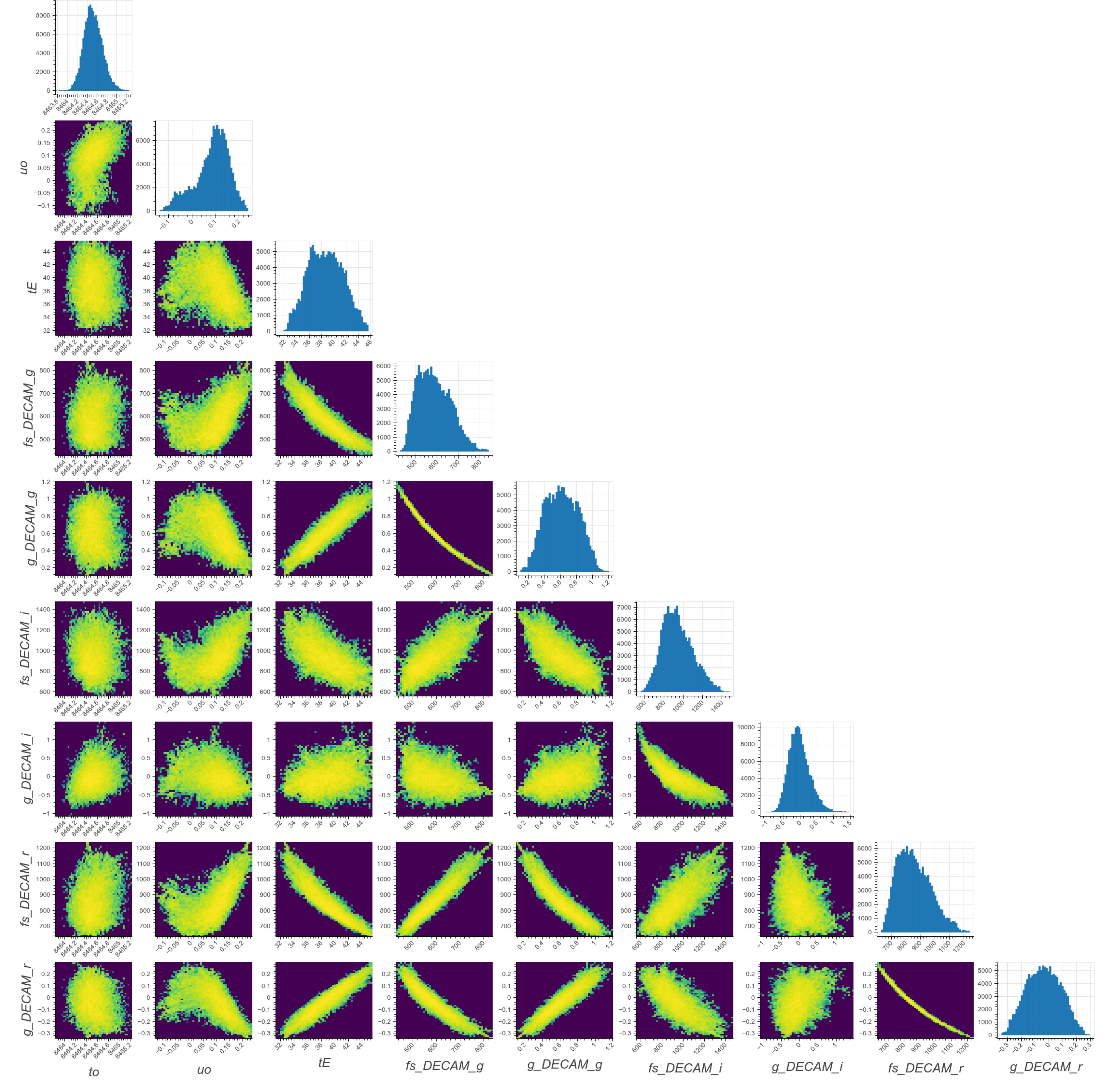}
    \caption{Corner plot for the PSPL model.}
    \label{Fig:fit_distr1}
\end{figure}

\begin{figure}[h]
    \includegraphics[width=1.0\textwidth]{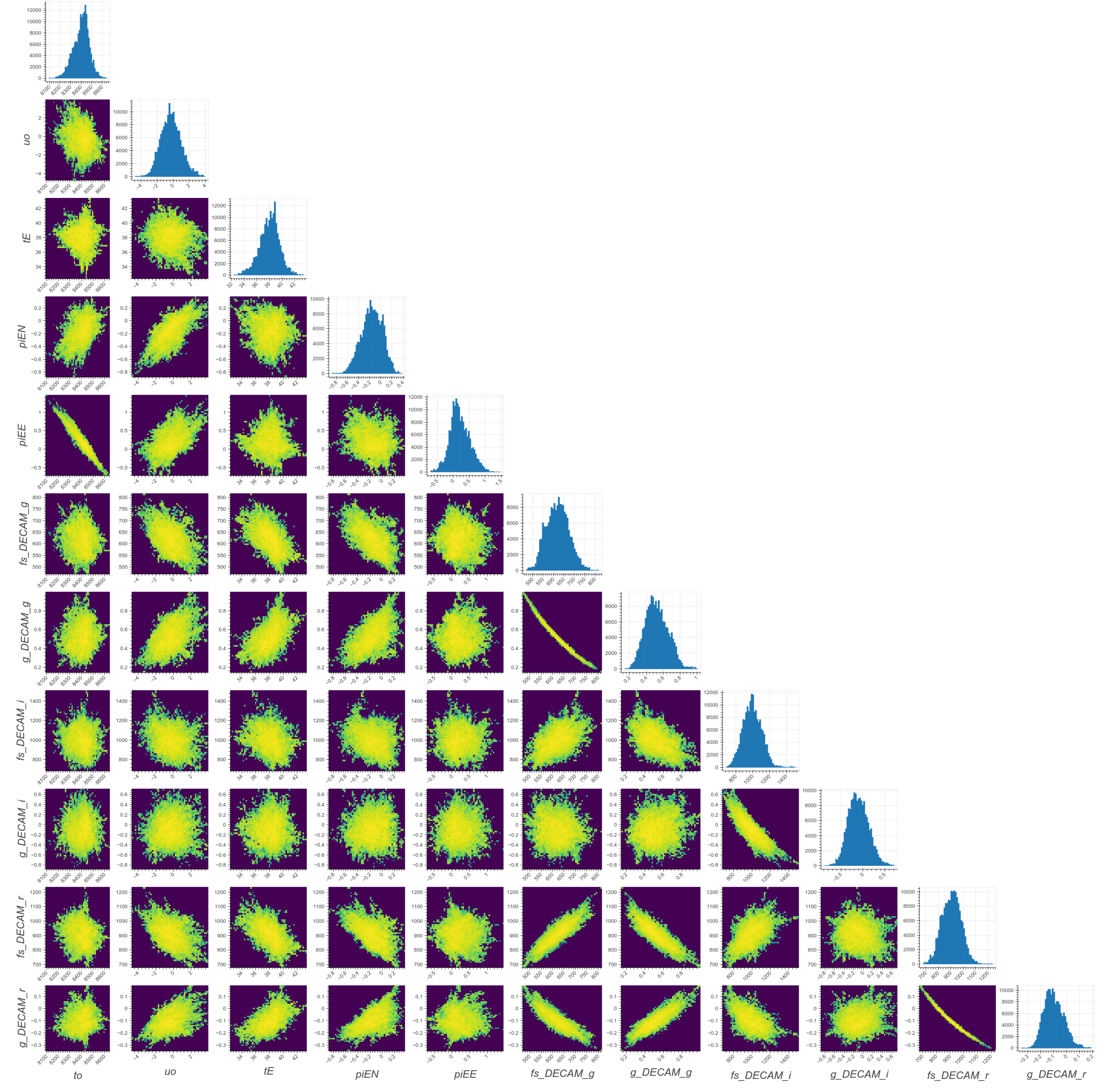}
    \caption{Corner plot for the PSPL + Earth Parallax model.}
    \label{Fig:fit_distr2}
\end{figure}

\begin{figure}[h]
    \includegraphics[width=1.0\textwidth]{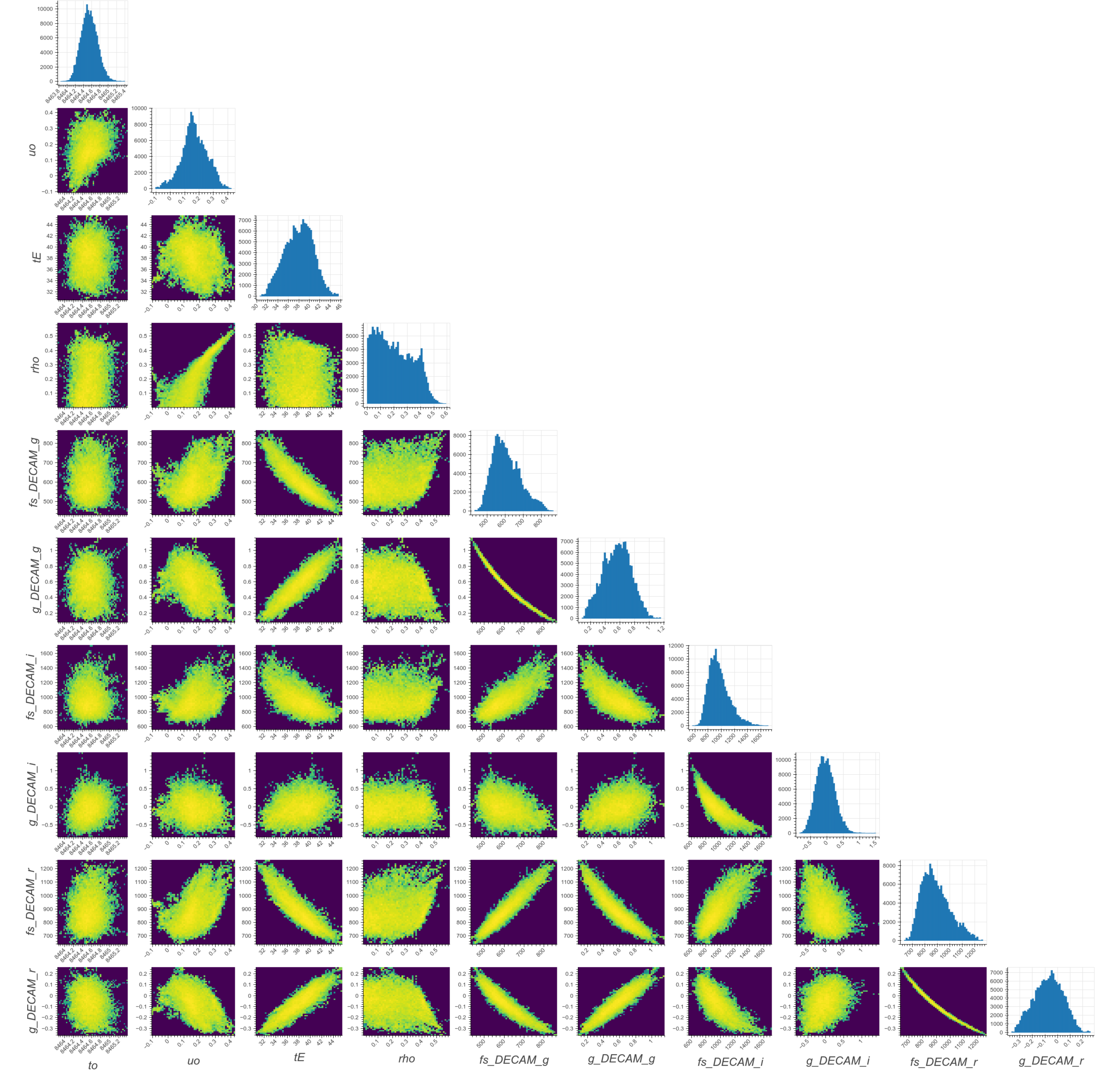}
    \caption{Corner plot for the FSPL model.}
    \label{Fig:fit_distr3}
\end{figure}

\begin{figure}[h]
    \includegraphics[width=1.0\textwidth]{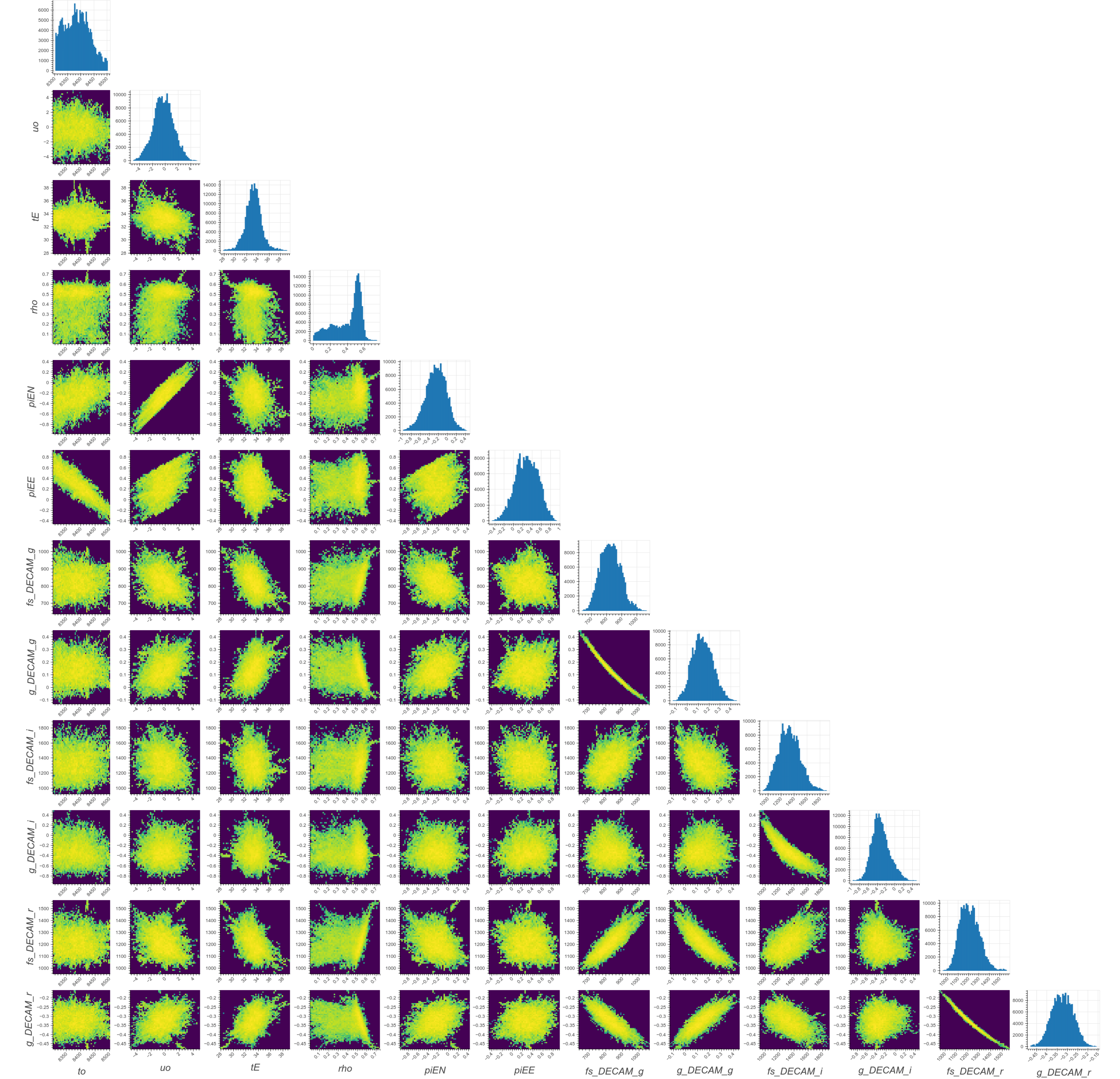}
    \caption{Corner plot for the FSPL + Earth Parallax model.}
    \label{Fig:fit_distr4}
\end{figure}

\end{document}